\documentclass[prb,twocolumn,superscriptaddress,floats,showpacs]{revtex4}

\usepackage[dvips]{graphicx}
\usepackage{amsmath}
\usepackage{booktabs}
\usepackage{dcolumn}

\begin{document}

\preprint{PGF-1}

\title{Low energy, quasi-one-dimensional, spin dynamics in charge-ordered
La$_{2-x}$Sr$_{x}$NiO$_4$}

\author{P. G. Freeman}
\email{freeman@ill.fr}\affiliation{Institut Laue-Langevin, BP 156, 38042 Grenoble Cedex
9, France}

\author{D. Prabhakaran}
\affiliation{Department of Physics, Oxford University, Oxford, OX1 3PU, United
Kingdom }

\author{K. Nakajima}
\affiliation{Neutron Science Section, MLF Division, J-PARC Center
2-4 Shirane Shirakata, Tokai, Naka, Ibaraki 319-1195, Japan}

\author{A. Stunault}
\affiliation{Institut Laue-Langevin, BP 156, 38042 Grenoble Cedex
9, France}
\author{M. Enderle}
\affiliation{Institut Laue-Langevin, BP 156, 38042 Grenoble Cedex
9, France}

\author{C. Niedermayer}
\affiliation{Laboratory for Neutron Scattering, ETHZ and PSI,
CH-5232 Villigen PSI, Switzerland}

\author{C. D. Frost}
\affiliation{ISIS Facility, Rutherford Appleton Laboratory,
Chilton, Didcot, OX11 0QX, United Kingdom}

%\author{H. Woo}
%\affiliation{Department of Physics, Brookhaven National
%laboratory, Upton, New York 11973, USA}

\author{K. Yamada}
\affiliation{ WPI Research Center, Advanced Institute for Materials Research, Tohoku University, Sendai 980-8577, Japan}

\author{A. T. Boothroyd}
\affiliation{Department of Physics, Oxford University, Oxford, OX1 3PU, United
Kingdom }

\date{\today}% It is always \today, today,

\begin{abstract} The low energy spin excitations of
La$_{2-x}$Sr$_{x}$NiO$_4$, $x = 0.275$ and $1/3$, have been
investigated by unpolarized- and polarized-inelastic neutron
scattering from single crystals. A pattern of magnetic diffuse
scattering is observed in both compositions, and is consistent with
quasi-one-dimensional AFM spin correlations along the charge
stripes. Analysis of the energy lineshape for $x = 1/3$ indicates
that the diffuse scattering is inelastic with a characteristic energy of
$1.40\pm0.07$\,meV. There is no discernible difference between the
diffuse scattering from $x = 0.275$ and $x = 1/3$, suggesting that
it is an intrinsic property of the charge stripes.

\end{abstract}

\pacs{75.40.Gb, 71.45.Lr, 75.30.Fv,75.30.Et}% PACS, the Physics and Astronomy
\maketitle

\section{\label{sec:intro}Introduction}

Striped patterns of spin and charge order have been observed in a
wide range of antiferromagnetic oxides since the initial discovery
in
La$_{1.48}$Nd$_{0.4}$Sr$_{0.12}$CuO$_{4}$.\cite{tranquada-Nature-1995}
Interest has been sustained by continuing uncertainty about the true
importance of stripe correlations for the mechanism of
superconductivity in the layered cuprates. In some models, for
example, stripe correlations assist in the formation of pairing
instabilities that can lead to superconductivity,\cite{mechanism}
whereas according to experiment static charge stripes suppress
superconductivity.\cite{tranquada-Nature-1995}. Investigations into
the fundamental properties of stripes are therefore potentially
important for an understanding of cuprate superconductivity, as well
as providing insight into an interesting emergent phase of
electronic matter.

Experimental investigations on stripe phases have been made on
various materials, but the layered nickelates
La$_{2-x}$Sr$_x$NiO$_{4+\delta}$ (LSNO) have been a particularly
informative model system to
study.\cite{neutron,x-ray,chen93,Yamada94,pash-PRL-2000,Yoshinari-PRL-2000}
LSNO, which is isostructural with the ``214" high temperature
superconductor La$_{2-x}$Sr$_x$CuO$_{4+\delta}$, exhibits spin and
charge stripe order for $0.15 \leq x \leq
0.5$.\cite{yoshizawa-PRB-2000,hatton-2002} The stripes form on the
square NiO$_2$ layers and consist of diagonal bands of
antiferromagnetically (AFM) ordered Ni$^{2+}$ spins separated by
charged domain walls that act as antiphase boundaries to the
magnetic order. At one-third doping (e.g.~$x = 1/3, \delta = 0$) the
stripe order is particularly stable owing to a combination of two
factors, firstly a commensurability effect that pins the charge
stripes to the lattice, and second a stripe periodicity which is the
same for the magnetic and a charge
order.\cite{ramirez-PRL-1996,yoshizawa-PRB-2000,kajimoto-PRB-2001}
%LSNO  does not  superconduct and only becomes metallic for $x> 0.9$.\cite{Sreedhar}
At this doping level LSNO displays long range ($>100$\,\AA) charge
order,\cite{yoshizawa-PRB-2000,hatton-2002} making it an ideal
material in which to probe the charge ordered state.
%Another LSNO doping level that is know to have a commensurate charge ordered state is the $x = 1/2$. The
%checkerboard charge ordered state of $x = 1/2$ is observed to have  dramatically enhanced ordering
%temperatures.\cite{chen-PRL-1993,freeman-PRB-2002,kajimoto-PRB-2003}

The magnetic excitation spectrum in the ordered stripe phase of LSNO
with $x \approx 1/3, \delta = 0$ has been investigated in some
detail by neutron inelastic scattering and found to contain two
distinct components, (i) quasi-two-dimensional spin-wave excitations
of the AFM-ordered regions, extending to $\sim$80\, meV in
energy,\cite{boothroyd-PRB-2003,bourges-PRL-2003,boothroyd-PhysicaB,Woo}
and (ii) a low energy ($<10$\,meV) quasi-one-dimensional (q-1D)
magnetic fluctuation consistent with short-range AFM correlations
along the charge stripes.\cite{boothroyd-PRL-2003} A two-component
spectrum with qualitatively similar characteristics has also been
observed in La$_{3/2}$Sr$_{1/2}$NiO$_4$.
(Ref.~\onlinecite{freeman-PRB-2005}) The observation of two types
of magnetic dynamics is consistent with the existence two magnetic
sub-systems associated with nominally Ni$^{2+}$ and Ni$^{3+}$ ions,
the latter of which form the charge stripes. At this time, however,
there is no microscopic model of the magnetic interactions in LSNO
that provides a unified description of the complete magnetic
spectrum.

The aim of the present study was to work towards a better
understanding of the low-energy q-1D magnetic correlations by
comparing neutron scattering data from two doping levels, $x=0.275$
and $x=1/3$ (both with $\delta=0$), and by examining in more detail
the energy lineshape of the magnetic scattering. We have found that
the q-1D scattering is present at both doping levels, and our
analysis of the energy lineshape reveals that the signal is gapped
at the minimum of the q-1D dispersion.

\section{\label{sec:exper}Experimental Details}

Single crystals of La$_{2-x}$Sr$_{x}$NiO$_{4}$ were grown by the
floating-zone method.\cite{Prabhakaran-JCG-2002} The crystals were
in the form of rods with typical dimensions 7--8\,mm in diameter and
$\sim$40\,mm in length (mass $\sim 15$\,g).

Neutron scattering measurements were performed on crystals with
$x=0.275$ and $x=1/3$ on the triple-axis spectrometers (TAS) IN8,
IN20 and IN14 at the Institut Laue-Langevin, and on RITA-II at SINQ,
Paul Scherrer Institut. The energies of the incident and scattered
neutrons were selected by Bragg reflection from crystal arrays of
pyrolytic graphite (PG) crystals (IN8, IN14, RITA-II), physically
bent Si crystals (IN8) or Heusler arrays (IN20).  The monochromators were vertically  focused (IN8, IN14) and horizontally focused (IN8, IN20, RITA-II)  to maximise neutron flux on the sample position.
The analyzers were horizontal focused on all instruments and
vertically focused on IN8.
%On IN8 and IN14 double focusing monochromators and anaylzers were used, and horizontal focusing %was used on %%@
%IN20 and RITA II.
Data were collected with fixed final neutron wavevectors of
2.662\,\AA$^{-1}$ (IN8, IN20), 1.5\,\AA$^{-1}$ (IN14, RITA-II) and
1.2\,\AA$^{-1}$ (IN14). A pyrolytic graphite filter (IN8, IN20) or
Be/BeO filter operating at 77\,K (IN14, RITA-II) was placed between
the sample and analyzer to suppress higher-order harmonic
scattering. On IN20 polarized neutrons were employed, and the
neutron spin polarization ${\bf P}$ was maintained in a specified
orientation with respect to the neutron scattering vector ${\bf Q}$
by an adjustable guide field of a few mT at the sample position. A monitor is placed between the monochromator and sample position to determine the number of  neutrons incident on the sample position. The monitor count has an energy dependent contamination due to higher order neutrons in the incident beam, which  we have corrected  for  when integrated intensities are shown. The crystals we aligned so that the horizontal scattering plane was
$(h,k,0)$ for $x = 1/3$ and $(h,h,l)$ for $x = 0.275$ (we refer here
to the tetragonal unit cell of the space group $I4/mmm$ with unit
cell parameters $a=3.8$\,{\AA} and $c=12.7$\,{\AA}). The particular
crystals used in the TAS measurements were grown at Oxford
University and have been used in previous neutron scattering studies
described
elsewhere.\cite{boothroyd-PRB-2003,Woo,freeman-PRB-2004,boothroyd-PhysicaB}

To supplement the TAS measurements we will also present some data
on the $x=0.275$ composition collected on the MAPS time-of-flight
spectrometer at the ISIS spallation neutron source.  A report on the findings of this study at higher energy transfers, can be found elsewhere.\cite{Woo} The sample used
on MAPS was an array of four crystals grown at Kyoto University and
co-aligned by X-ray diffraction to within about $1^{\circ}$.
Crystals with $x=0.275$ from the same source have been used in
neutron diffraction studies of magnetic and charge
order.\cite{lee-PRB-2001} The sample was mounted on MAPS in a
closed-cycle refrigerator and aligned with the $c$ axis parallel to
the incident beam direction. A Fermi chopper was used to select the
incident neutron energy of 60\,meV. The intensity was normalized and
converted to units of scattering cross-section
(mb\,sr$^{-1}$\,meV$^{-1}$\,[f.u.]$^{-1}$) by comparison with
measurements from a standard vanadium sample. Scattered neutrons
were recorded in large banks of position-sensitive detectors. The
spin dispersion in La$_{2-x}$Sr$_{x}$NiO$_{4}$ is highly
two-dimensional,\cite{Woo} and so we project the data onto the $(h,
k)$ two-dimensional reciprocal lattice plane. The elastic energy
resolution on MAPS was 2.7\,meV (full-width at half-maximum).

\section{\label{sec:results} Results}

\begin{figure}[!t] \begin{center}
\includegraphics[clip=, width=8cm]{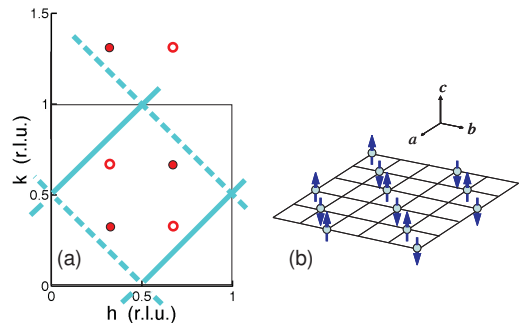} \caption[q-1D scattering]{(color online)
Diagrams representing the low-energy magnetic fluctuations in
La$_{5/3}$Sr$_{1/3}$NiO$_4$, after
Ref.~\onlinecite{boothroyd-PRL-2003}. (a) Schematic map of the low
energy scattering in the $(h,k,0)$ plane of reciprocal space.
Circles are stripe superlattice zone centres and diagonal lines
indicate the diffuse scattering from ideal one-dimensional AFM spin
chains running parallel to the charge stripes. Filled circles and
full lines are for stripes running parallel to the
$[1,\overline{1},0]$ direction; open circles and broken lines are
for stripes parallel to the $[1,1,0]$ direction. (b) Model of AFM
correlations along the charge stripes consistent with the observed
q-1D diffuse scattering. The arrows indicate the instantaneous
orientation of the spins on the charge stripes. For clarity, the
background AFM order in the regions between the charge stripes is
not shown.} \label{fig:previous}
\end{center} \end{figure}

To help visualise the experimental measurements we reproduce in
Fig.~\ref{fig:previous} part of a figure from our previous
publication on the q-1D magnetic fluctuations in
La$_{5/3}$Sr$_{1/3}$NiO$_4$ (Ref.~\onlinecite{boothroyd-PRL-2003}).
Figure~\ref{fig:previous}(a) is a simplified map of the low-energy
scattering features in the $(h,k,0)$
plane of reciprocal space.  The sharp peaks (circles) at $(1/2\pm1/6, 1/2\pm1/6, 0)$ and $(1/2\pm1/6, 
1/2\mp1/6, 0)$ represent spin-wave scattering associated with the magnetic ordering wavevectors. There are two 
pairs of peaks because there are two possible orientations of the stripes on the NiO$_2$ layers, either along 
the $[1,\overline{1},0]$ direction or the $[1,1,0]$ direction. The diagonal lines which run parallel to the 
stripe directions represent the approximate pattern of diffuse scattering observed in 
La$_{5/3}$Sr$_{1/3}$NiO$_4$, Ref.~\onlinecite{boothroyd-PRL-2003}. Figure~\ref{fig:previous}(b) depicts an 
array of AFM chains running parallel to the stripes which would give rise to the diagonal grid of diffuse 
scattering shown in Fig.~\ref{fig:previous}(a). In reality, the the line of the observed diffuse scattering is 
not exactly straight but meanders slightly so as to follow approximately the magnetic zone boundaries, 
suggesting that the fluctuations on adjacent chains are weakly correlated. The spins are shown pointing upwards to reflect the observation that the strength of the out-of-plane fluctuations is about twice the strength of 
the in-plane fluctuations.\cite{boothroyd-PRL-2003} The diffuse scattering has a strong dynamic component on a
THz frequency scale, and may even be entirely dynamic. Whether the q-1D scattering is quasielastic or inelastic
(i.e. gapped) is an open question which will be addressed later in this work.

\begin{figure}[!ht] \begin{center}
\includegraphics[width=8cm,clip=]{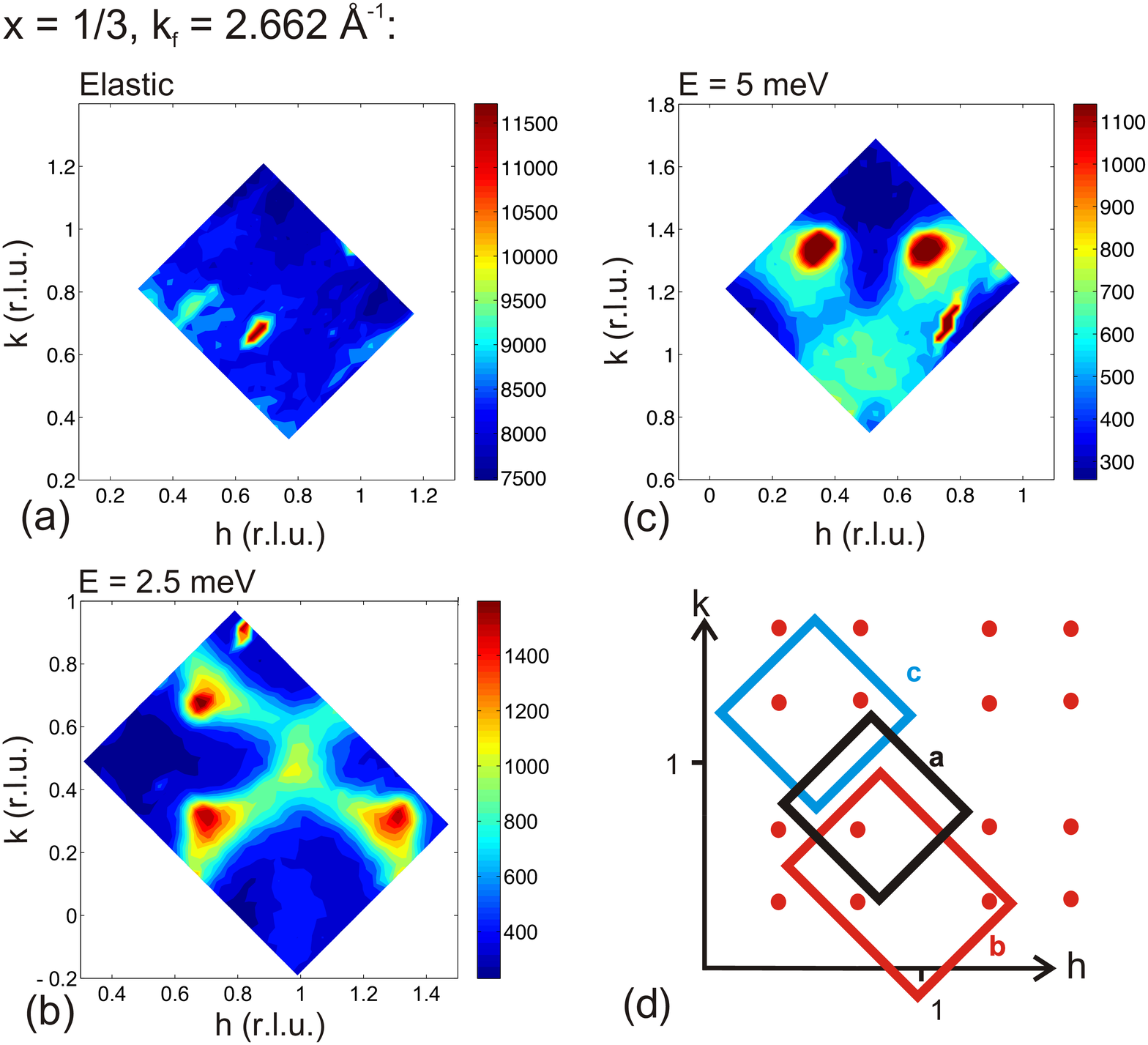}
\caption[Low energy excitations of x = 0.333]{(Color online) Maps of the scattering
intensity in the $(h, k, 0)$ reciprocal space plane of LSNO $(x = 1/3)$. (a) Elastic scattering, (b) 
$E=2.5$\,meV, and (c) $E=5$\,meV. The data were collected on IN8 with a final wavevector of $k_f = 
2.662$\,{\AA}$^{-1}$ at a temperature of 2\,K. The contour plots are formed from $\sim$25$\times 30$ 
rectangular grids of data points.
%The intensity is capped at 70,000\cite{harmonics} in (a) and
%1800 counts in (c) to emphasise the weaker scattering.
The data in (a) and (c) were measured with a pyrolytic graphite monochromator, and (b) with a silicon 
monochromator. (d) Diagram of reciprocal space indicating the areas mapped out in (a)--(c).} 
\label{fig:x=033E=0+5meV}
\end{center}
\end{figure}

Figure~\ref{fig:x=033E=0+5meV} presents maps of the magnetic scattering intensity measured in the $(h, k, 0)$ 
plane of the $x =1/3$ crystal at a temperature of 2\,K. The maps were collected at three fixed energies, (a)
0\,meV (i.e.~elastic scattering within the resolution of the spectrometer), (b) 2.5\,meV, and (c) 5\,meV. The 2.5\,meV data is reproduced from Ref.~\onlinecite{boothroyd-PRL-2003}. Figure~\ref{fig:x=033E=0+5meV}(d) shows the areas covered in the three maps, which are not the same.

In the elastic map, Fig.~\ref{fig:x=033E=0+5meV}(a), there is a magnetic Bragg reflection at $(0.667, 0.667, 
0)$ due to the pattern of AFM order between the charge stripes, and a small spurious peak at $(0.45,0.75,0)$.
No diffuse elastic scattering signal can be observed within the experimental precision. Comparing the two 
inelastic maps, Figs.~\ref{fig:x=033E=0+5meV}(b) and \ref{fig:x=033E=0+5meV}(c), one can see that the q-1D 
diffuse scattering is broader at $E = 5$\,meV than at at $E = 2.5$\,meV. This is consistent with our previous 
measurements\cite{boothroyd-PRL-2003} which showed that the q-1D scattering disperses in the direction 
perpendicular to the scattering ridge, with a bandwidth of about 10\,meV.  The intensity of the q-1D scattering 
is modulated, with maxima adjacent to the spin-wave scattering from the AFM order and at the positions where 
the diffuse ridges meet, i.e.~$(1, 0.5, 0)$ and equivalent
positions.

\begin{figure}[!ht] \begin{center}
\includegraphics[width=8cm,clip=]{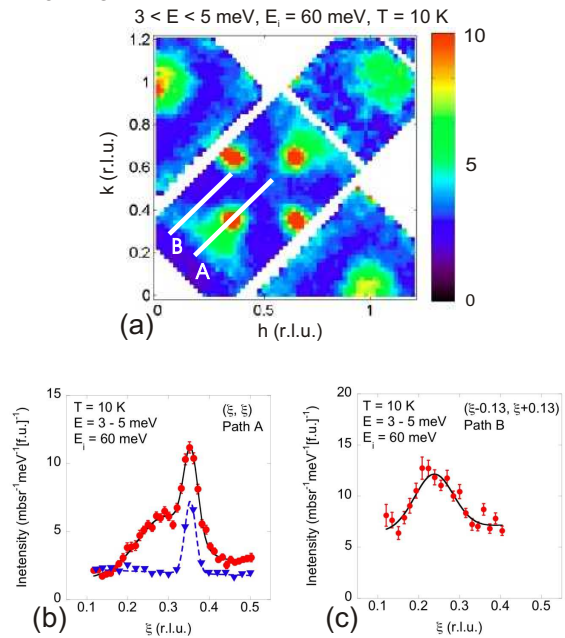}
\caption[Low energy excitations of x = 0.275]{(Color online) Neutron scattering from La$_{2-x}$Sr$_{x}$NiO$_4$ 
($x=0.275$) measured on MAPS. (a) Slice through the data volume averaged over the energy range 3--5\,meV. (b) 
and (c): cuts along the paths marked A and B in (a).  An elastic scattering cut along path A is also displayed 
in (b). Solid and dashed lines indicate fits to the data with either one or two gaussian peaks on a sloping 
background.
%We note that the elastic intensity of the magnetic order may contribute to Path A, due to the energy being %%@
%less than two resolution FWHM from the elastic position.
%and dashed lines in (b) are fits to the data of respectively two and one gaussian peak shapes on a
%sloping background, the solid line in (c) is a fit to the data of a single gaussian peak shapes on a
%sloping background.
} \label{fig:x=0_275E=3-5meV}
\end{center}
\end{figure}

We now turn to the LSNO crystal with $x = 0.275$. At this composition the charge stripe order is incommensurate 
with the crystal lattice and it of interest to see whether this has any effect on the q-1D magnetic 
fluctuations.  Figure \ref{fig:x=0_275E=3-5meV}(a) shows the distribution of scattering intensity from LSNO 
($x=0.275$) measured on MAPS. The scattering has been averaged over the energy range 3--5\,meV and plotted as a 
function of the in-plane components $(h, k)$ of reciprocal space. There are four strong scattering signals at 
positions $(0.5 \pm \epsilon /2 , 0.5 \pm  \epsilon /2)$ and $(0.5 \pm \epsilon /2 , 0.5 \mp  \epsilon /2)$ 
with  $\epsilon = 0.297 \pm 0.001$, from the steeply dispersing spin-wave excitations of the AFM order. In Fig  \ref{fig:x=0_275E=3-5meV}(a)  the spin wave excitations from the AFM order are too low in energy to be  resolved into spin wave cones, and appear simply as spots.  In addition to these, there can 
also be seen ridges of diffuse scattering similar to the q-1D scattering observed from the $x = 1/3$ crystal 
[Fig.\ \ref{fig:x=033E=0+5meV}(b,c)]. Figures
\ref{fig:x=0_275E=3-5meV}(b) and \ref{fig:x=0_275E=3-5meV}(c) show
cuts through this data along the paths marked A and B in Fig.\
\ref{fig:x=0_275E=3-5meV}(a).  Path B is chosen so  that  no intensity from the excitations of the ordered AFM is observed in a scan along path B. The location for path B is determined experimentally, by performing scans perpendicular to path A through the excitations from the AFM order.
The cut in
Fig.~\ref{fig:x=0_275E=3-5meV}(b) along path A shows
the diffuse scattering as a shoulder to the sharper spin-wave peak, whereas the cut shown in Fig.\ 
\ref{fig:x=0_275E=3-5meV}(c) along path B shows just the diffuse scattering peak, which
is centred on $\xi \approx 0.25$ [the $\xi$ coordinate measures the position along the scan projected 
perpendicularly onto line A, i.e.\ such that $\xi=0$ projects onto $(0,0)$ and $\xi=0.5$ projects onto $(0.5,0.5)$]. 
From these and similar cuts we find that the diffuse scattering at
$x =0.275$ follows the same slightly-meandering path as does the diffuse scattering at $x = 1/3$ --- see Figs.\ 
\ref{fig:x=033E=0+5meV}(b) and \ref{fig:x=033E=0+5meV}(c). In Fig. \ref{fig:x=0_275E=3-5meV}(b) we also display 
the elastic scattering intensity along path A. This shows the magnetic Bragg peak from the AFM order but does 
not contain any elastic signal corresponding to the position of the diffuse inelastic signal.

\begin{figure}[!ht]
\begin{center}
\includegraphics[width=8cm,clip=]{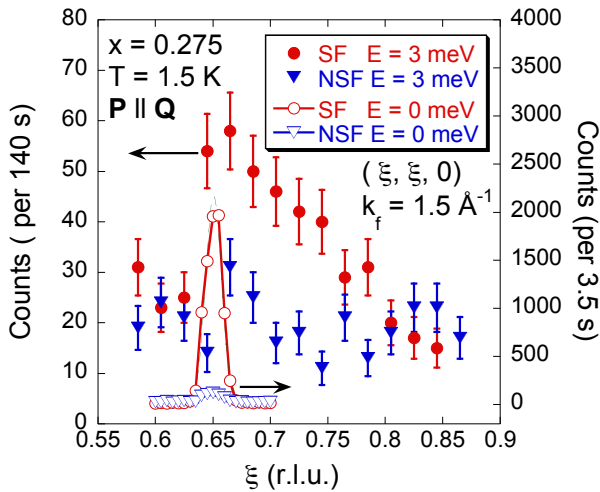}
\caption[Polarized neutron scattering Q scan of the low energy
excitations of x = 0.275]{(Color online) Polarized neutron
scattering from La$_{2-x}$Sr$_{x}$NiO$_4$ ($x = 0.275$). The scans
are along the $(1, 1, 0)$ direction [path A in Fig.\
\ref{fig:x=0_275E=3-5meV}(a)] and measured at energies of 0\,meV
(elastic scattering, right scale) and 3\,meV (left scale). Spin-flip
(SF) and non-spin-flip (NSF) scattering channels are shown. The
small peak in the NSF channel of the elastic scan is due to
imperfect polarization, which has not been corrected for.}
\label{fig:Polx=0_275.eps}
\end{center}
\end{figure}

To confirm that the diffuse scattering signal is magnetic in origin
we performed scattering measurements on IN20 employing neutron
polarization analysis. For these measurements we constrained the
neutron polarization ${\bf P}$ to be parallel to the scattering
vector ${\bf Q}$. In this configuration scattering from electronic
magnetic moments causes the neutron spin to flip, whereas scattering
via non-magnetic processes does not.

Figure \ref{fig:Polx=0_275.eps} displays the neutron spin-flip (SF)
and non-spin-flip (NSF) scattering from the $x = 0.275$ crystal
along a line equivalent to path A. Plots of elastic scattering data
and inelastic scattering data with $E = 3$\,meV are shown. The
elastic scan contains a strong peak in the SF channel centred on
$\xi = 0.65$, the AFM ordering wavevector for this composition. A
small peak in the NSF channel at the same position is due to
imperfect spin polarization, which has not been corrected for. At $E
= 3$\,meV there is a broad peak centred on $\xi \approx 0.7$ in the
SF channel but not in the NSF channel. Due to their very steep dispersion, the
spin-wave scattering from the  AFM order only accounts for the delta-shaped left-side of the SF peak.\cite{Woo}
The remaining extent of the SF peak can be accounted for by the inelastic
diffuse scattering.

The results of the unpolarized and polarized neutron scattering
measurements presented here for $x = 0.275$ are qualitatively very
similar to our previous measurements\cite{boothroyd-PRL-2003} on $x
= 1/3$. This shows that the q-1D diffuse scattering is magnetic in
origin and has the same character in both the $x = 0.275$ and $x =
1/3$ samples.

\begin{figure}[!ht] \begin{center}
\includegraphics[,clip=, width = 8 cm]{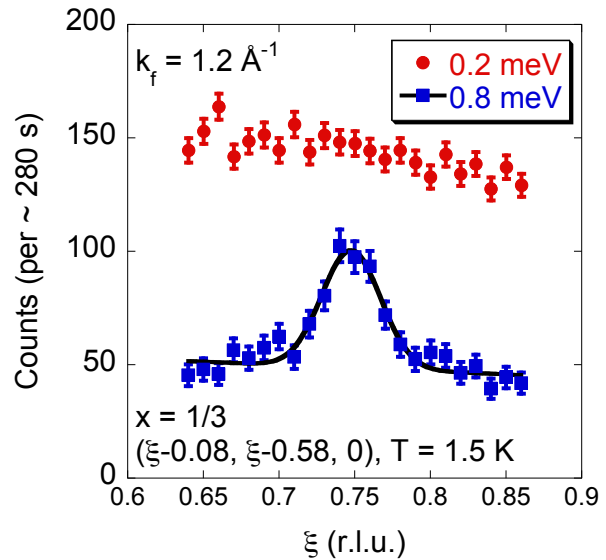} \caption[Evidence
of a spin gap in x = 1/3]{(Color online) Constant-energy scans along
path B for the La$_{2-x}$Sr$_{x}$NiO$_4$ ($x = 1/3$) at 0.2\,meV and
0.8\,meV.  The solid line is the best fit to a gaussian function on
a linear background. } \label{fig:x=0.333E=0.2,0.8}
\end{center}
\end{figure}

Next we turn to the question of whether the q-1D scattering is
gapped or not. To answer this we performed inelastic neutron
scattering measurements on IN14 on the $x = 1/3$ crystal at energies
$E \leq 3$\,meV. Figure \ref{fig:x=0.333E=0.2,0.8} shows
constant-energy scans at 0.2\,meV and 0.8\,meV  along a direction
equivalent to path B of Fig.\ \ref{fig:previous}(a). The scan at
0.8\,meV reveals a peak centred on $\xi = 0.75$ from the q-1D
fluctuation, whereas the scan at 0.2\,meV shows no peak at $\xi \sim
0.75$ within the statistical precision of the data. The peak widths
of both the q-1D diffuse scattering at 0.8\,meV and the spin-wave
scattering at 0.6\,meV (not shown) are
resolution-limited\cite{restrax}.

\begin{figure}[!ht] \begin{center}
\includegraphics[width=8cm,clip=]{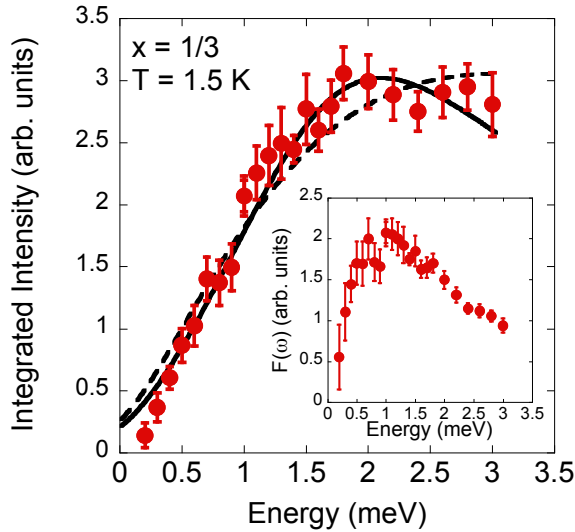} \caption[Evidence of
a spin gap in x = 1/3 part2 ]{(Color online) Lineshape of the q-1D
diffuse scattering in La$_{2-x}$Sr$_{x}$NiO$_4$ ($x = 1/3$) at low
energy. The integrated intensities were obtained from the area of a
Gaussian function fitted to a series of constant-energy scans like
those shown in Fig.\ \ref{fig:x=0.333E=0.2,0.8}. For $E \leq
0.6$\,meV the signal was small and the Gaussian width was fixed in
the fit to the value obtained from the $0.8$\,meV data. The
measurements were made with $k_{\rm f} = 1.2$\,{\AA}$^{-1}$ for $E
\leq 1$\,meV and $k_f = 1.5$\,\AA$^{-1}$ for $E \geq 1$\,meV, and
normalized so as to match at $1$\,meV. The solid line is calculated
from Eq.\ (\ref{Gap}) with $E_0 = 1.4$\,meV, and the broken line is
calculated with $E_0 = 0$. The inset shows the spectral weight
function, which is related to the dynamical susceptibility by $\chi''(\omega) \propto \omega
F(\omega)$.} \label{fig:in14x=0.333} \end{center} \end{figure}

\begin{figure}[!ht] \begin{center}
\includegraphics[clip=,width=8cm]{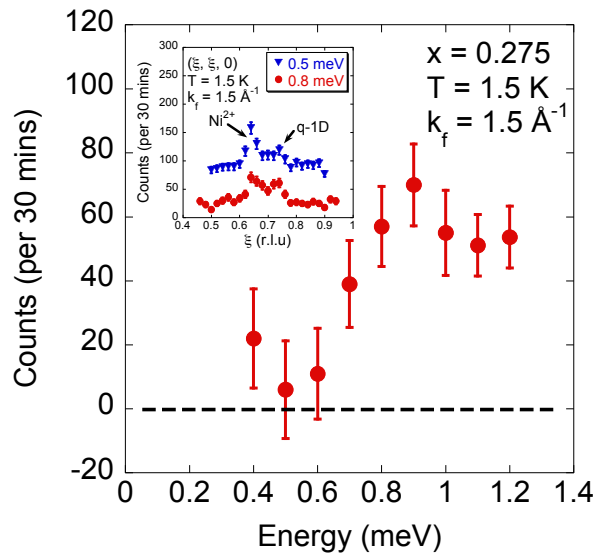}
\caption[Evidence of a spin gap in x = 0.275]{(Color online)
Background-corrected energy scan of the q-1D diffuse scattering from
La$_{2-x}$Sr$_{x}$NiO$_4$ ($x = 0.275$) at ${\bf Q} = (0.75, 0.75,
0)$. The background was estimated from energy scans at $(0.75, 0.75,
0) \pm (0.19, 0.19, 0)$. Inset: Constant-energy scans along path A
at energies of 0.5\,meV and 0.8\,meV. Arrows indicate the positions
of the spin-wave scattering from the AFM order of the Ni$^{2+}$
spins and from the q-1D diffuse scattering. The 0.8\,meV scan has
been offset by the addition of 50 counts for clarity.}
\label{fig:x=0_275lowE}
\end{center} \end{figure}

By performing a series of constant-energy scans  like those of Fig.\
\ref{fig:x=0.333E=0.2,0.8} and fitting these with gaussian peaks we
determined the integrated intensity of the spin-wave and q-1D
diffuse scattering peaks as a function of energy. We remind the reader  that in scans along path B only the q-1D excitations are observed, while in scans along path A the excitations from the AFM order are clearly resolved from the q-1D.\cite{boothroyd-PRL-2003} Below 3\,meV we
observed no energy variation of the integrated intensity of the
spin excitations from the AFM order along path A(not shown), placing an upper limit of 0.3\,meV on
the in-plane anisotropy gap. In figure \ref{fig:in14x=0.333} the
energy variation of the integrated intensity of the diffuse
scattering peak in $x = 1/3$ is plotted. With increasing energy
transfer the integrated intensity of the diffuse scattering
increases monotonically up to $\sim 2$\,meV, then remains almost
constant up to 3\,meV, the highest energy measured.

In a separate experiment  we investigated the q-1D diffuse scattering from the $x = 0.275$
sample, performed on the RITA-II spectrometer.  In the inset of Figure \ref{fig:x=0_275lowE} we show constant-energy
scans  at 0.8\,meV and 0.5\,meV along a direction equivalent to path
A. For $E = 0.8$\,meV  the q-1D scattering is clearly observed and
centred on $\xi = 0.73$. At $E = 0.5$\,meV there is still a small
excess of scattering above the background at $\xi = 0.73$. The centring is consistent with the meandering of the q-1D observed in  Fig. \ref{fig:x=033E=0+5meV}, and the energy dependence is 
consistent with our observations of the q-1D in the $x =1/3$ sample, see Fig.\ref{fig:x=0.333E=0.2,0.8}  and  Fig.\ref{fig:in14x=0.333}. Figure\ \ref{fig:x=0_275lowE} shows a constant-${\bf Q}$ scan at ${\bf Q}=(0.75,0.75,0)$ of the background-corrected amplitude of the q-1D in the $x = 0.275$, off-centred to avoid the excitations from the AFM order.
% To remove any possibility of measuring excitations from the AFM order, we performed an off centred %constant-${\bf Q}$ scan of the q-1D at ${\bf Q}=(0.75,0.75,0$ on the $x = 0.275$ sample. 
 The background was estimated from measurements at nearby wavevectors. With decreasing energy transfer the signal remains roughly constant down to 0.8\,meV, then drops to a level
close to zero at around 0.5\,meV. Within the limitations of  low counting statistics,  this is consistent with the costant energy scans on the $x = 0.275$, %sample displayed in the inset to Fig. \ref{fig:x=0_275lowE}, 
and our observations of the q-1D in the $x =1/3$ sample, see Fig.\ref{fig:in14x=0.333}.

\section{\label{sec:Discussion}Discussion}

This work has been concerned with the nature of the low-energy q-1D
diffuse scattering in La$_{2-x}$Sr$_{x}$NiO$_4$ first found for $x =
1/3$ (Ref.\ \onlinecite{boothroyd-PRL-2003}). We have shown here
that the diffuse scattering is also present at $x = 0.275$, and that
it is the same for $x = 0.275$ as for $x = 1/3$ to within the
experimental precision. This implies that the q-1D scattering is not
dependent on the periodicity of the spin-charge stripe order, and
neither is it a consequence of the special circumstance found at
$x=1/3$ in which the spin and charge order have the same periodicity
and are commensurate with the crystal lattice. Instead, the diffuse
scattering appears to be an intrinsic property of individual charge
stripes embedded in the AFM matrix formed by the Ni$^{2+}$ spins,
consistent with the interpretation in terms of q-1D correlations
among spins in the charge stripes.\cite{boothroyd-PRL-2003}

In this work we have also carefully measured the low-energy
lineshape of the q-1D diffuse scattering. To model the lineshape we
recall that the dynamical part of the scattering is proportional to
the linear response function
\begin{equation}
S({\bf Q},\omega) = \frac{1}{\pi} \{n(\omega) + 1\} \chi''({\bf
Q},\omega), \label{Gap}
\end{equation}
where $\hbar \omega$ is the energy transferred to the system,
$\chi''({\bf Q},\omega)$ is the imaginary part of the dynamical
susceptibility, and
\begin{equation}
n(\omega)=\frac{1}{\exp(\beta\hbar\omega)-1}.\label{eq:Planck}
\end{equation}
To describe the {\bf Q}-integrated intensities we used the
phenomenological Lorentzian lineshape constructed to satisfy
Detailed Balance,
\begin{equation}
\chi''(\omega)=\left[\frac{\Gamma\omega/\pi}{(\hbar\omega
-E_0)^{2}+\Gamma^{2}}+\frac{\Gamma\omega/\pi}{(\hbar\omega
+E_0)^{2}+\Gamma^{2}}\right] \label{Gap1}\end{equation} where
$\Gamma$ is the Lorentzian width (half-width at half-maximum) and
$E_0$ is the energy of the undamped mode. As the q-1D has a large intrinsic energy width, $E_0$ is the characteristic energy for the q-1D, not a gap energy.

The best fit to the data achieved by this lineshape is shown as the
solid line in Fig.\ \ref{fig:in14x=0.333}, and gives an energy $E_0
= 1.40\pm0.07$\,meV and $\Gamma = 1.57\pm 0.13$\,meV.  The fact that
the fitted $E_0$ is non-zero provides a clear indication that the
q-1D diffuse scattering is inelastic rather than quasielastic. To
assess how robust this result is we repeated the fit with $E_0$
fixed at zero, corresponding to a quasielastic lineshape. The best
fit thus obtained is shown with a broken line in Fig.\
\ref{fig:in14x=0.333}. The width of the quasielastic fit was $\Gamma
= 3.0\pm0.3$\,meV. The fit to the inelastic lineshape has a
goodness-of-fit parameter $\chi^{2}$ = 1.19 compared with a
$\chi^{2}$ = 2.25 for the quasielastic lineshape. On the strength of
this evidence we conclude that q-1D diffuse scattering corresponds
to a gapped inelastic excitation.

For an alternative representation of the inelastic lineshape we show
in the inset to Fig.\ \ref{fig:in14x=0.333} the spectral weight
function $F(\omega)$, which is related to the imaginary part of the
dynamical susceptibility by $\chi''(\omega) \propto \omega
F(\omega)$. The large energy width of the excitations indicates that
the q-1D excitations are relatively short-lived, $\Delta t \sim
\hbar/\Gamma \sim 4 \times10^{-13}$\,s.

In our original work on the q-1D spin correlations in LSNO with $x =
1/3$ we analyzed the magnetic spectrum with respect to that of an
antiferromagnetic spin chain.\cite{boothroyd-PRL-2003} Assuming the
doped holes reside in localized Ni$^{3+}$ states with low-spin
$S=1/2$ this implies that the fundamental excitations are spinons
with a gapless dispersion. In the present study, however, we have
presented evidence that the spectrum has a gap of 1.4\,meV. Further
thought is needed, therefore, if we are to reconcile this
information with our understanding of the q-1D magnetic diffuse
scattering.

The model of an AFM spin chain to describe the spin correlations
along the  Ni$^{3+}$ charge stripes is reasonable providing the
coupling between the spin chain and the AFM order of the Ni$^{2+}$
spins can be neglected. The justification for neglect of this
coupling is that the net Heisenberg exchange acting on the Ni$^{3+}$
sites from the AFM-ordered Ni$^{2+}$ spins cancels at the mean-field
level, so the coupling between the spin chain and the AFM order is
frustrated. However, if the individual Ni$^{3+}$--Ni$^{2+}$ exchange
interactions are strong enough then a weak ferromagnetic (FM) order
could be induced on the Ni$^{3+}$ stripes combined with a canting of
the AFM order of the Ni$^{2+}$ spins. Such a possibility has been
proposed by Klingeler {\it et al.} based on magnetization
data.\cite{Klingeler-PRB-2005} Weak FM order would be difficult to
detect in diffraction experiments because the associated magnetic
Bragg peaks would coincide with the Bragg peaks from the crystal
lattice. Nevertheless, the competition between induced FM order and
AFM correlations from an effective AFM exchange along the stripes
could provide an explanation for the observed gapped dispersion.
Such a model has been investigated recently with promising
results.\cite{long}

\section{\label{sec:Conclusion}Conclusion}

This work has revealed that gapped, quasi-1D AFM spin correlations
are an intrinsic property of the charge-stripes in
La$_{2-x}$Sr$_{x}$NiO$_4$. The findings should inform theoretical
models for the magnetic interactions in LSNO, and hence contribute
to a broader understanding of the formation and stability of
spin--charge stripes in LSNO and related systems.\\

\section{\label{sec:Acknowledgments}Acknowledgments}

The authors would  like to acknowledge the help of H. J. Woo in the MAPS experiment. This work was performed in part at the Swiss Spallation Neutron
Source SINQ, at the Paul Scherrer Institute (PSI), Villigen,
Switzerland. We are grateful for support from  the Engineering and
Physical Sciences Research Council of Great Britain,  the
European Commission under the 7th Framework Programme through the
`Research Infrastructures' action of the `Capacities' Programme,
Contract No: CP-CSA INFRA-2008-1.1.1 Number 226507-NMI3, and  the support of  a Grant-in-Aid for Scientific Research (No. 22244039) from the MEXT, Japan.

\end{document}